\documentstyle[epsfig,12pt]{article}
\newcommand{\definmath}[2] {\def#1{\ifmmode#2\else$#2$\fi}}
\definmath{\GeV}  {\mathrm{GeV}}
\definmath{\GeVc} {{\mathrm{GeV}}\!/c}
\definmath{\GeVcc}   {{\mathrm{GeV}}\!/c^2}
\def\Journal#1#2#3#4{{#1} {#2} (#4) #3 }

\def\PLB{{ Phys. Lett.}  B}

\begin{document}

\begin{titlepage}

\begin{flushright}
{\bf GLAS-PPE/1999--16}\\
{\bf November 1999}\\
\end{flushright}
\vspace{-1.0cm}
\begin{center}
\epsfig{file=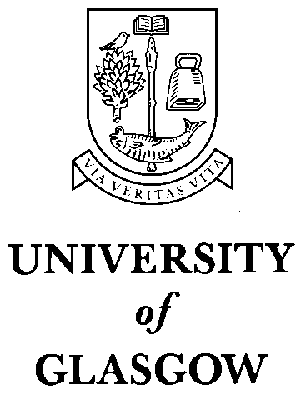,width=3.0cm}
\end{center}
\vskip 1cm
\begin{center}
\vskip 1cm
{\large\bf Precision Measurements of Particle}\\ 
{\large\bf Masses using Jets at LEP2}
\vskip 1cm
{\bf J.~Jason~Ward}\\
{\em Department of Physics and Astronomy}\\
{\em University of Glasgow}\\
{\em Glasgow, G12 8QQ}\\
{\em United Kingdom}\\
\vskip 1.5cm
{\em E-mail: Jason.Ward@cern.ch}\\
\vskip .5cm
\end{center}

\vskip 1cm

\begin{abstract}
How massive elementary particles get their mass is one of the greatest
open questions in physics.  Two of the major goals of the current LEP2
physics program that help to address this question are (a) to measure as 
precisely as possible the W Boson mass $m_{\rm W}$ and (b) to exclude or 
discover the Higgs Boson within the available kinematic region.  The 
reconstruction of invariant masses with jets from 4-jet channels (e.g. 
$WW \rightarrow q{\bar q}q{\bar q}$ and HZ $\rightarrow b{\bar b}q{\bar q}$) 
and missing energy channels (e.g. $WW \rightarrow e{\bar \nu}q{\bar q}$ and 
$e^{+}e^{-} \rightarrow H\nu{\bar \nu}$) is discussed.  The emphasis is on 
the determination of $m_{\rm W}$, from which the r{\^o}le of calorimetry 
in such a precision measurement is emphasised.
\end{abstract}
\vspace{1.0cm}
\begin{center}
{\em Invited Talk at the VIII International Conference on Calorimetry in}\\
{\em High Energy Physics, Lisbon, Portugal, June 13-19, 1999.}
\end{center}

\end{titlepage}
\newpage
\pagestyle{plain}
\setcounter{page}{1}

\section{Introduction}
How elementary particles obtain the property of mass is one of the greatest 
unanswered questions in physics.  Measuring as precisely as possible the masses 
of elementary particles, particularly gauge bosons, will be vital data for
the theoretical physicist wanting to tackle the mass mechanism problem.
Then when a candidate theory exists, such as the Higgs mechanism, it has to be 
tested by experiment.
    
In the first phase of the LEP program (LEP1), the mass and width of the 
Z boson were measured to be $m_{\rm Z} = 91.1867 \pm$ 0.0021 GeV and 
$\Gamma_{\rm Z} = 2.4939 \pm$ 0.0024 GeV~\cite{lepEWwg99-15}, via 
measurements of {\it cross-sections} around the Z-peak.  

In the second phase of the LEP program (LEP2), pairs of W bosons are produced.  
The new experimental challenge is to measure $m_{\rm W}$ by {\it direct 
reconstruction} of the W decay products with a precision that matches that 
of the {\it indirect} measurements of $m_{\rm W}$ ($\sim 30$ MeV).
Significant disagreement between the direct and indirect determinations of 
$m_{\rm W}$ might indicate the breakdown of the Standard Model.  As the 
$e^+e^-$ centre-of-mass energy continues to increase, we also 
search for a Higgs boson signal. 
%look at our data to see if the Higgs boson has lightly reared its head.

\section{W Mass by Direct Reconstruction at LEP2}\label{subsec:wmass}

\subsection{W Pair Production}\label{subsec:wpp}
At LEP2, the three basic W-pair production diagrams contain  
$Z \rightarrow WW$, $\gamma \rightarrow WW$ and t-channel 
neutrino exchange (which dominates near the W-pair threshold).  
The issue of directly reconstructing the mass of a heavy boson 
naturally focusses on these events, since W-pairs have already been 
produced in their thousands in each of the four LEP detectors.  
Typical statistics of W pair events are shown in Table~\ref{tab:nums}.   

\begin{table}[h]
\caption{Year, $e^+e^-$ centre-of-mass energy, integrated luminosity
per LEP experiment and approximate number of W-pair events per LEP 
experiment.\label{tab:nums}}
\begin{center}
\footnotesize
\begin{tabular}{|c|c|c|c|}
\hline
Year &\raisebox{0pt}[13pt][7pt]{$\sqrt{s}$ (GeV)} &
\raisebox{0pt}[13pt][7pt]{$\mathcal{L}$(pb$^{-1}$)/expt} & {$N_{\rm WW}$/expt}\\
\hline
1996 & 161 & $\sim$ 10  & $\sim$ 35   \\ 
     & 172 & $\sim$ 10  & $\sim$ 120  \\ 
1997 & 183 & $\sim$ 57  & $\sim$ 850  \\ 
1998 & 189 & $\sim$ 175 & $\sim$ 2700 \\ 
\hline
\end{tabular}
\end{center}
\end{table}

Given that BR(W $\rightarrow q\overline{q}$) $\simeq$ 68\% and that 
the remaining decays are leptonic, one sees that the W pair sample subdivides 
as follows:  hadronic channel (WW $\rightarrow$ $q\overline{q}q\overline{q}$) 
$\simeq 46\%$, semileptonic channels (WW $\rightarrow$ $l\nu q\overline{q}$) 
$\simeq 44\%$ and the fully leptonic channels (WW $\rightarrow$ $l\nu l\nu$) 
are $\simeq 10\%$, where $l$ denotes $e,\mu$ or $\tau$. 

\subsection{Selecting WW Events}\label{subsec:SeleWW}
The r{\^o}le of calorimetry in characterising WW events is implicitly 
assumed in this brief outline of selection.  Hadronic WW events are 
characterised by four hadronic jets, and the total missing energy 
and momentum are small.  Typical preselection of hadronic events 
would use information on missing energy, multiplicity,
spericity and thrust.  A final selection would be based on kinematical 
variables or a multidimensional analysis (e.g. neural network).
Selections are highly efficient ($\sim 85\%$) but notably 
not entirely pure (purity $\sim 80\%$) with the background mostly 
coming from $q\overline{q}(\gamma)$ events.

The semileptonic events are characterised by two hadronic jets, 
one isolated high momentum lepton and large missing momentum 
(using both direction and magnitude information).  
Choosing the lepton within the event uses lepton identification,
in addition to the fact that, at LEP energies not too far above 
the W-pair threshold, the charged lepton is likely to be the track 
with the highest momentum component antiparallel to the missing 
momentum.  $q\overline{q}(\gamma)$ events and 4-fermion events 
are most of the background, but the selection purity, typically
80-95\%, is greater than in the hadronic channel.

Fully leptonic WW events have large missing energy and missing 
$p_T$ and two acoplanar, acollinear leptons.  Selection of these events 
uses lepton identification and event topology.  Since there are two 
neutrinos, one from each W, extracting the W mass from these events 
relies on using the lepton energy spectra.  This non-jet channel is 
not discussed, although there is a recent ALEPH result for $m_{\rm W}$
determined using the fully leptonic channel~\cite{ALEPHmwlep,Imma}.

\subsection{Jets, Leptons and Kinematic Fitting}
In a selected W-pair event with two or four jets, the reconstructed W 
mass is obtained from the di-jet mass.  One sees, assuming the simple  
case of massless jets, that the di-jet mass $m_{J_1 J_2}$ depends
on the jet {\it energies} and {\it angles};

\vspace{-0.1cm}
\[ m_{J_1 J_2}^2 \simeq 2 E_{J_1} E_{J_2} (1-\cos \theta_{J_1 J_2}).\]

Jets are obtained by clustering detector objects with a chosen algorithm.  
%The choice of algorithm can take into account certain 
%considerations, e.g. aiming to reduce particle mixing of 
%decay products of two different W's in the hadronic channel.
%A selected hadronic event is usually forced into four jets, 
%introducing the problem of how to decide which pair of jets belong 
%to which parent W (there are 3 combinations for four-jet events). 
The hadronic channel poses particular problems arising from mis-assignment
of soft particles to the wrong W, interconnection effects between particles 
from different W's, such as Bose-Einstein correlations and colour-reconnection,
and jet combinatorics (three ways to form two di-jets from four-jet events).
In the semileptonic channels, calorimeter objects created near 
the charged lepton, by Final State Radiation or Bremsstrahlung,
can evade association to the lepton and thus degrade the 
event mass estimator.

Table~\ref{tab:resolutions} compares jet and lepton 
resolutions for two LEP experiments, OPAL~\cite{OPALmw183} and 
ALEPH~\cite{ALEPHperf}.  One should note that the energy of 
an electron is measured much more precisely than a jet.
The intrinsic energy resolution for an OPAL lead glass block is 
considerably better than the quoted resolution for 47 GeV electrons
in OPAL implies, for a number of reasons~\cite{MThomson}.
Predominantly this is because there are $\sim 2 X_0$ of material 
in front of the lead glass blocks, mostly the magnet coil and 
pressure vessel.  Smaller effects come from the 
reduction of the gain of the photomultipliers at LEP2, which was 
done to increase their dynamic range to allow for 100 GeV
electrons, and the fact that the calibration of the barrel 
region is more difficult with a smaller sample of 
central detector Bhabhas.

\begin{table}[t]
\caption{Examples of jet and lepton resolutions from OPAL 
and ALEPH.\label{tab:resolutions}}
\begin{center}
\footnotesize
\begin{tabular}{|c|c|}
\hline
\raisebox{0pt}[13pt][7pt]{OPAL} &
\raisebox{0pt}[13pt][7pt]{ALEPH} \\
\hline
 & \\ 
%Tracking (Jet Chamber): 
Combined tracking: 
%& Tracking (TPC) \\
& Combined tracking: \\
%$\sigma_{\phi}\sim 0.3$ mrad, $\sigma_{\theta}\sim 1$ mrad. & \\
$(\sigma_{p_T}/p_T)^2 = (0.020)^2 + (0.0015 p_T)^2/{\rm GeV}^2$ & 
$\sigma_{p_T}/p_T = 0.0006 p_T + 0.005$ \\
& \\ 
Electromagnetic Calorimeter: 
& Electromagnetic Calorimeter: \\
11704 lead glass blocks.  For Barrel: 
& Lead/wire plane sampling device \\
$\sigma_E/E\simeq 0.2\% + 6.3\%/\sqrt{E (GeV)}$ 
& $\sigma_E/E \simeq 0.9\% + 18\%/\sqrt{E (GeV)}$ \\
For  $E_{elec} \simeq$ 47 GeV, $\Delta E/E \simeq$ 3\%. 
& For $E_{elec} \simeq$ 47 GeV, $\Delta E/E \simeq$ 3.5\%.\\ 
& \\ 
Jet Energy Resolution (Z Peak):
& Jet Energy Resolution (Z Peak): \\
$\sigma_E/E \simeq 20\%$ 
& $\sigma_E/E \simeq 11\%$ \\
& \\
Jet Angular Resolution:
& Jet Angular Resolution: \\
20-30 mrad 
& 20-30 mrad \\
(depending on $E_{jet}$ and $\theta_{jet}$)
& (depending on $E_{jet}$ and $\theta_{jet}$) \\
& \\
\hline
\end{tabular}
\end{center}
\end{table}

The next stage of analysis is the kinematic fit.  The aim of such a 
fit is to evaluate a set of four 4-vectors (two for each W) for each 
event, consistent with (a) $(E,p)$ conservation
(the LEP beam energy is known to high precision, $\Delta E_{beam} 
\simeq 20$ MeV for $\sqrt{s}=189$ GeV data); (b) Expected biases 
(determined from WW MC and detector simulation) e.g. average jet energy 
loss for a given $(E_{jet},\theta_{jet})$; (c) Measured energies and 
angles of the jets and leptons, within their resolutions.  

Using $(E,p)$ conservation in the hadronic channel imposes 
four constraints (4C) and results in two mass values per event
(for a given di-jet combination).  However, an additional 
requirement that these two masses are equal may be imposed, resulting 
in a total of five constraints (5C) and only one mass value per event.
This is made possible by the fact that the mass resolution of the 
reconstruction, which is about 2-3 GeV, is very similar to the W width,
$\Gamma_{\rm W}$.  In the semileptonic channels, three constraints are 
lost in reconstructing the neutrino, thus 4C and 5C reduce to 1C and 2C 
respectively.

Kinematic fitting results in the significant improvement of 
mass resolution.  The choice of using the additional equal mass constraint 
varies between LEP experiments.  One can appreciate at this stage that 
since a tight constraint on energy is being provided by the 
beam energy, the most important information from jets is their 
direction.

For small angle jets, detector simulation predicts that the 
reconstructed jet energy is 30-40\% lower than the true jet energy.
This is not surprising, since as the polar angle decreases (towards 
the beamline) there are numerous subdetector boundaries and changes, 
and more emphasis on electromagnetic calorimetry for luminosity
measurements.  The expected energy loss is implemented in the 
kinematic fit, however one must ensure it correctly reproduces 
the effect in data.  This is achieved by looking at 
$Z\rightarrow q\overline{q}$ events at the Z-pole, noting that in a 
two-jet event, the true jet energy should be the same value as the 
precisely known beam energy.  Thus any discrepency in energy loss 
between data and Monte Carlo can be accounted for, and for the LEP 
experiments it is typically 2-3\% in the most forward regions.
DELPHI additionally use 3-jet events~\cite{DELPHImw183}.  ALEPH 
correct for the effect, using the error on the correction as a 
systematic error~\cite{ALEPHmw183}.

Three of the four LEP experiments use a Monte Carlo reweighting 
method~\cite{OPALmw183,ALEPHmw183,L3mw99}.  A large sample of Monte Carlo 
events generated with a known $m_{\rm W}$ are reweighted to different 
$m_{\rm W}$ values using matrix element information, until the MC mass 
distributions best fit the data.  If the MC correctly models the data, 
biases are implicitly accounted for.  ISR is implemented in the MC up 
to $\mathcal{O}(\alpha^2 L^2)$ and a systematic error can be estimated to 
represent the omission of higher order terms.  DELPHI employ a convolution 
method~\cite{DELPHImw183}.  
   
\subsection{Example of Calorimeter Uncertainties (ALEPH)}

A number of methods are used to calibrate the electromagnetic 
calorimeter (ECAL) of ALEPH~\cite{ALEPHperf}.  The ECAL gain is 
directly monitored by looking at an Fe$^{55}$ source 
(which ages~\cite{Callot}).  The amplitude 
of the gain variation is $\sim 2-3\%$ over one year, which 
after correction is stable to better than 0.3\%.

For a range of electron energies, the ratio of ECAL energies 
to electron track momenta can be measured from various processes
at the Z-peak: $e^+e^- \rightarrow e^+e^-e^+e^-$ yields electrons 
in the 1-10 GeV range and $Z \rightarrow \tau^+\tau^- 
(\tau \rightarrow e\nu\overline{\nu})$ electrons in the 10-30 GeV
range.  Also the $Z \rightarrow e^+e^-$ and Bhabha processes
produce electrons at the beam energy (45.6 GeV).

For high energy LEP runs, one can measure $E_{ECAL}/E_{beam}$ 
for Bhabha events. With large data samples, the high energy 
runs can also be split into smaller runs to estimate the time 
dependence of ECAL variations.

The typical net ECAL energy calibration uncertainty is 
$\pm (0.7-0.9)\%$.

The hadronic calorimeter (HCAL) of ALEPH also uses physics 
processes for calibration~\cite{ALEPHperf}.  The idea is to 
constrain the peak 
of the muon energy distribution in HCAL to its expected 
position ($\sim$ 3.7 GeV for muons crossing the calorimeter) which 
is measured in beam test.  Then at the start of every data-taking period, 
$Z \rightarrow \mu^+\mu^-$ and $Z \rightarrow q\overline{q}$ events
are used to calibrate.  The use of hadronic Z decays provides 
a much more statistically powerful sample, which can be used 
because the ratio between the average energy released by hadronic 
Z decays and an isolated muon in an HCAL module is well 
known from data.  This technique gives a `time 0' uncertainty 
of $\pm$ 1\%.

For high energy running it is possible to compare data and MC
for \mbox{$\gamma\gamma\rightarrow\mu^+\mu^-$ events}, yielding muons 
in the energy range 2.5-10 GeV.  The energy distributions agree 
at the 1.5\% level.  

The typical net uncertainty is thus $\pm$ 2\%.

The effects of the calorimeter uncertainties on the $m_{\rm W}$
measurements are evaluated by changing, in Monte Carlo samples,
the calibration of the subdetectors by the net 
uncertainties described above, and performing mass fits 
before and after such a change.
    
\subsection{Using $Z\gamma$ Events}
The emission of a hard ISR $\gamma$ reduces the effective 
centre-of-mass energy of $e^{+}e^{-}$ interactions, such that 
the effective interaction can be at the Z resonance.  
Kinematic reconstruction of these so-called ``$Z\gamma$'' events,
with $Z\rightarrow q\overline{q}$, can provide a cross-check to 
W mass measurements, by measuring either $m_{\rm Z}$ or $E_{beam}$,
and comparing these values to independent measurements.

L3 measure $m_{\rm Z}$ from $q\overline{q}\gamma$ events~\cite{L3mw99}
at $\sqrt{s}= 189$ GeV using $\sim$ 10K events.  This is a check 
of detector calibration, jet reconstruction, fitting method etc., 
and uses a kinematic fit and a reweighting technique
as in their W mass analysis.  The fitted value from $Z\gamma$ events is 
$m_{\rm Z}$ = 91.106 $\pm$ 0.062 GeV (prel.), in good agreement with 
$m_{\rm Z}$ extracted from their cross-section measurements at the Z-pole 
of $m_{\rm Z}$ = 91.195 $\pm$ 0.009 GeV.  This represents an important 
test of the complete mass analysis method.  

Since the beam energy is used in the kinematic fits of WW events, 
the uncertainty on it propagates through to the W mass measurement.
The error on the beam energy is very precise at LEP1, less than 1 MeV, as 
it is measured by resonant depolarisation.  Polarisation has not been 
achieved above a beam energy of 60 GeV, so the same technique does not 
work at LEP2.  Instead, beam energy measurements at lower beam energies
are extrapolated to higher energies, resulting in a larger beam energy 
error (25 MeV at $\sqrt{s}=183$ GeV~\cite{lepewg}).  As a cross-check of the 
LEP determination of $E_{CM}$, ALEPH use the $q\overline{q}\gamma$ events 
by providing the precise ($m_{\rm Z}, \Gamma_{\rm Z}$) information
and using the jet angles to extract $E_{CM}$.  The average LEP 
centre-of-mass energy at ALEPH for the $\sqrt{s}=183$ GeV run of 1997
is measured to be 182.50 $\pm$ 0.19(stat) $\pm$ 0.08(syst) GeV~\cite{ALEPHzgamma}, 
which is consistent with the estimate from the LEP energy working 
group~\cite{lepewg} of 182.652 $\pm$ 0.050 GeV.     

\subsection{$m_{\rm W}$ Results}
Preliminary ALEPH results from the 189 GeV data are shown in 
Table~\ref{tab:ALEPH189res}.  This example illustrates the size 
of calorimetric-based systematic errors (i.e. calibration uncertainties 
and jet corrections, shown in bold) relative to other systematic 
errors.  One should remember that these systematic errors are evaluated 
with finite MC and data samples, and are therefore subject to some 
degree of statistical fluctuation.  Combining these results with data 
from previous years, and with the measurement of $m_{\rm W}$ from the 
lepton energy spectrum at 183 GeV~\cite{ALEPHmwlep}, ALEPH obtain the 
following preliminary result~\cite{ALEPHmw189}:

\vspace{-0.1cm}
\[ m_{\mathrm W} =\ \ 80.411
\pm 0.064{\mathrm{\small (stat.)}} 
\pm 0.037{\mathrm{\small (syst.)}}
\pm 0.022{\mathrm{\small (theory)}}
\pm 0.018{\mathrm{\small (LEP)}}
~{\rm GeV}\]

which has a total error of $\sim 80$ MeV.  The LEP combined  
result for the measurement of $m_{\rm W}$ by direct reconstruction, 
based on preliminary results available at the time of the Spring 1999 
conferences~\cite{Imma}, was $m_{\rm W} = 80.368 \pm 0.065$ GeV.
When combined with the W mass derived from WW cross-section measurements 
at and above threshold, this becomes $m_{\rm W} = 80.370 \pm 0.063$ GeV.
The error is dominated by the systematic error. 

\begin{table}[t]
\caption{
Summary of the correlated and uncorrelated systematic errors on $m_{\rm W}$.
The results are ALEPH data (preliminary) from the 189 GeV run.
$\ddagger$ denotes error taken from 183 GeV studies.\label{tab:ALEPH189res}}
\begin{center}
%\footnotesize
\begin{tabular}{|l|c|c|c|c|}\hline
           & \multicolumn{4}{c|}{$\Delta m_{\rm W}$ (MeV)} \\
\hline
Error source   & 4q & ${\mathrm e}$ & $\mu$ &$\tau$\\
\hline
Statistical          & 116  & 180   & 164   & 332 \\
\hline
 {\it Correlated Systematics:}          &      &       &       &       \\
 Fragmentation           & 35$\ddagger$   &  25$\ddagger$   &  25$\ddagger$   &  30$\ddagger$   \\
 {\bf Calorimeter calibrations}    & {\bf 30}   &  {\bf 27}   &  {\bf 14}   &  {\bf 19}   \\
 Tracking                          &  -   &   7   &   3   &   3   \\
 {\bf Jet corrections}             &  {\bf 8}   &  {\bf 14}   &   {\bf 4}   &   {\bf 7}   \\
 Initial state radiation & 10$\ddagger$   &   5$\ddagger$   &   5$\ddagger$   &   5$\ddagger$   \\
 LEP energy                        & 17   &  17   &  17   &  17   \\
\hline
 {\it Uncorrelated Systematics:}        &      &       &       &       \\
 Reference MC Statistics           & 10   &  16   &  15   &  23   \\
 Background contamination          & 10$\ddagger$   &   8   &   1   &  25   \\
 Colour reconnection               & 25$\ddagger$   &   -   &   -   &   -   \\
 Bose-Einstein effects             & 50$\ddagger$   &   -   &   -   &   -   \\
\hline
Total Systematics                  & 77   &  47   &  37   &  53   \\
\hline
\end{tabular}
\end{center}
\label{tab:syst-summary}
\end{table}

\section{Standard Model Higgs Boson}

Production of the Standard Model Higgs boson at LEP2 would mainly 
proceed via the `Higgsstrahlung' process ($Z^*\rightarrow HZ$). 
Additional small contributions would come from ZZ or WW fusion processes
(resulting in He$^+$e$^-$ and H$\nu\overline{\nu}$ final states 
respectively).  

\begin{figure}[htbp]
%\begin{flushleft}
\begin{tabular}[t]{@{}ll@{}}
\mbox{\epsfig{file=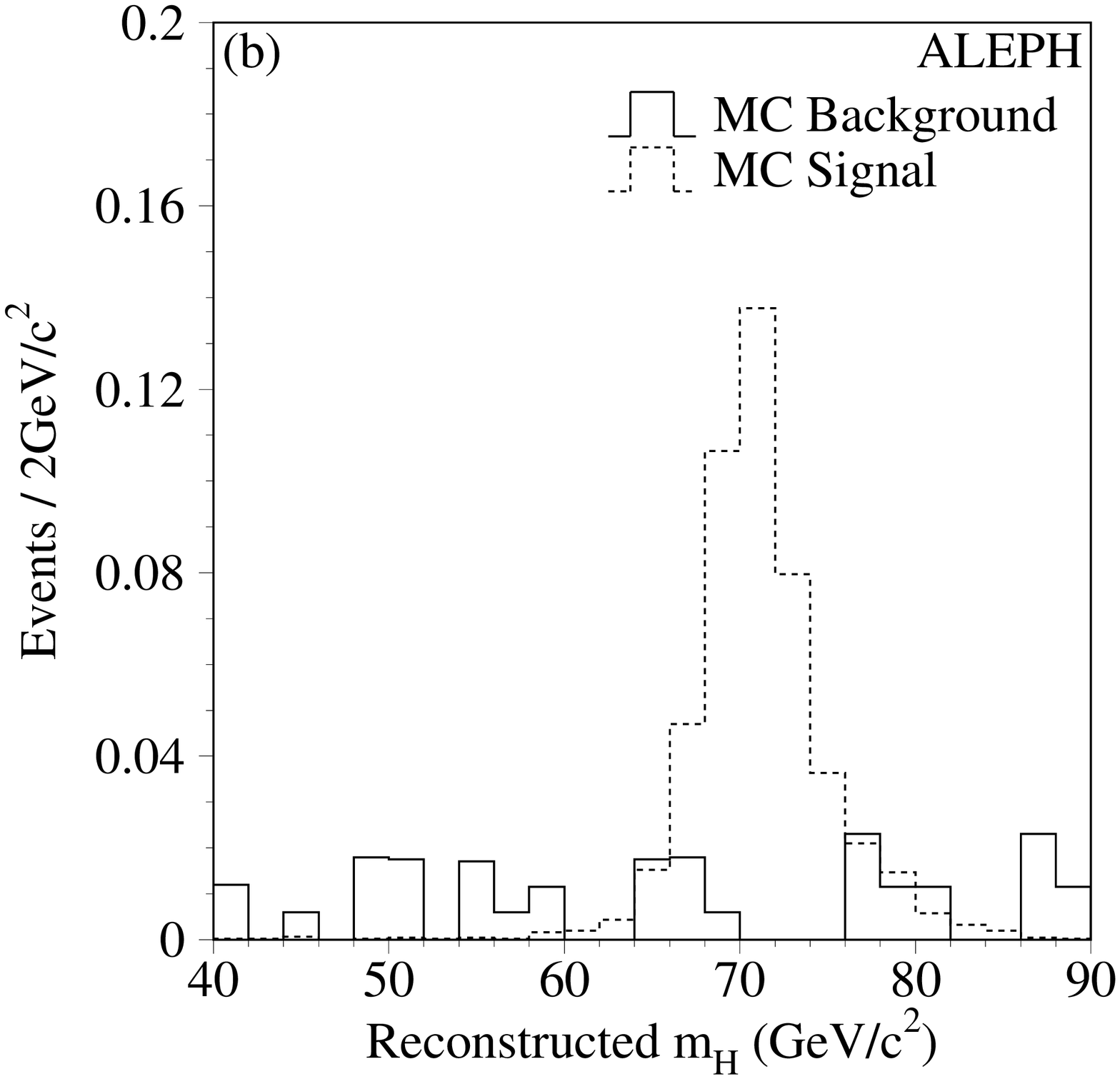,height=7.0cm}}
&
\mbox{\epsfig{file=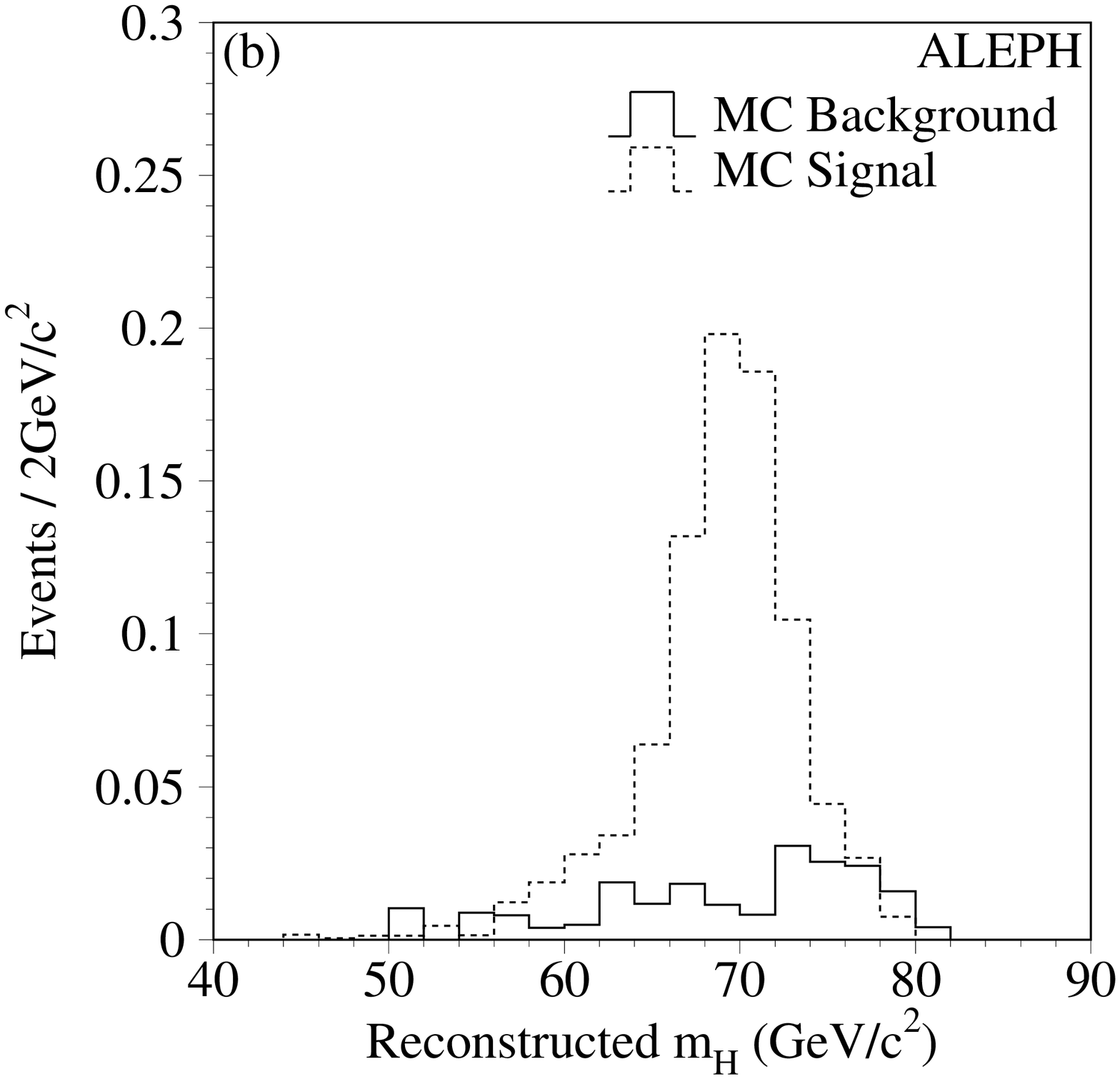,height=7.0cm}}
\end{tabular}
%\end{flushleft}
\caption{Left: distribution of the mass recoiling in the He$^+$e$^-$
and H$\mu^+\mu^-$ channels after all selection criteria are applied.  
Right: The distribution of the reconstructed Higgs boson mass
in the H$\nu\overline{\nu}$ channel.  The solid histograms are 
background and the dashed histograms are signal.}
\label{fig:higgsmass1}
\end{figure}

\begin{figure}[htbp]
\epsfxsize=15pc % will enlarge or reduce the postscript figures based on the xsize
\begin{center}
\epsfbox{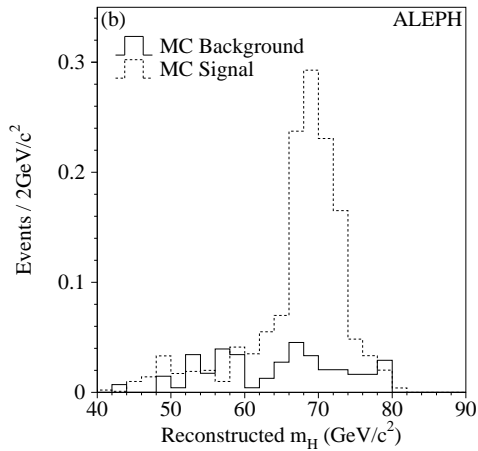} % postscript image file name
\end{center}
\caption{Distribution of the reconstructed Higgs boson mass in the 
channel HZ$\rightarrow$ b$\overline{\rm b}$q$\overline{\rm q}$.
The solid histogram is background and the dashed histogram is signal.
\label{fig:higgsmass2}}
\end{figure}

Using all of the available Electroweak measurements at the time of 
the Spring 1999 conferences, the Higgs mass was indirectly 
determined~\cite{Felcini} to be $m_{\rm H} = 71^{+75}_{-42} \pm 5$ 
GeV (central value and 68\% C.L. errors) and $m_{\rm H} < 220$ GeV 
at 95\% C.L.  For data collected at $\sqrt{s} \leq 183$ GeV, the 
combined LEP direct search lower mass limit is $m_{\rm H} > 89.7$ GeV.  
The reconstructed $m_{\rm H}$ in candidate events, from all final 
states, is a variable entering the calculation of confidence levels.
Figures~\ref{fig:higgsmass1} and~\ref{fig:higgsmass2} show MC Higgs mass 
distributions ($m_{\rm H} = 70$ GeV) for various final states, taken from 
an ALEPH analysis~\cite{ALEPHhiggs}.

The H$l^+l^-$ ($l = e$ or $\mu$) final state represents 6.7\% of the 
Higgsstrahlung cross-section.  It is characterised by two oppositely 
charged isolated leptons, where the mass of the lepton pair is 
close to the Z mass.  The recoil mass can be calculated from the 
$ll(\gamma)$ system, where the $\gamma$ refers to any final state 
radiation that is emitted.  Due to the high resolution of charged 
leptons (see Table~\ref{tab:resolutions}), the reconstructed Higgs 
mass also has high resolution. 

The missing energy channel H$\nu\overline{\nu}$ represents 20\%
of the Higgsstrahlung cross-section.  In the event of a signal, 
there would be large missing energy, 
a missing mass near the Z mass, acoplanar jets and the possibility
of b-tagging these jets.  The Higgs mass would have to be reconstructed 
from the di-jet mass, so it has larger width and a longer low-mass
tail in the reconstructed mass distribution compared to that of 
the H$l^+l^-$ channel.  This is clearly apparent in 
Figure~\ref{fig:higgsmass1}.  Forward calorimetry that is as 
hermetic as possible plays an important part in the event selection,
contributing to the rejection of $Z\gamma$, $We\nu$ and $Zee$ events.      

The four-jet channel, HZ$\rightarrow$ b$\overline{\rm b}$q$\overline{\rm q}$,
comprises 64.6\% of the Higgsstrahlung cross-section.
A signal in this final state would contain four isolated jets.  
A kinematic fit can be employed, 
after which the key trick is to plot $m_{12}+m_{34}-m_{Z}$.
The subscripts refer to jet pairings options.  In Figure~\ref{fig:higgsmass2}
the subsequent low mass tail due to jet combinatorics can be seen, 
which exists even though only the combination for which 
$m_{12}$ is closest to the nominal Z mass is shown.

\section{Summary}
This contribution has discussed in some detail the direct reconstruction 
of $m_{\rm W}$ at LEP.  Even though detector performance is intricately 
linked to complex analytical procedures and varying features of
different channels, it is clear that calorimetry plays a central   
r{\^o}le in the $m_{\rm W}$ measurement.  A repeated feature is 
the use of precisely-known quantites ($m_{\rm Z}$ and LEP beam energy)
to improve errors on mass measurements.  As LEP data continues to accumulate,
understanding systematic errors, including detector calibration uncertainties 
and possible jet angular biases, become even more important.

\section*{Acknowledgments}
I would like to thank the organisers of the conference for arranging 
an excellent meeting.  Prof. D.~Saxon and the University of Glasgow 
are gratefully acknowledged for funding my participation.  Many ALEPH 
colleagues have helped in the preparation of this contribution, particularly 
A.~Blondel, R.~Tenchini, F.~Ligabue, P.~Teixeira-Dias and E.~Lan\c{c}on.
Thanks also to M.~Thomson (OPAL) and S.~Gentile (L3) for their fast 
response to questions.

\end{document}